\documentclass[a4paper]{jpconf}
\usepackage{graphicx}

\begin{document}
\title{The apparent decay of pulsar magnetic fields}

\author{A Biryukov$^{1,2}$, A Astashenok$^{3}$, S Karpov$^{4,2}$ and G Beskin$^{4,2}$}

\address{$^{1}$ Sternberg Astron. Inst. of MSU, 13 Universitetsky pr., Moscow 119234, Russia}
\address{$^{2}$ Kazan Federal University, 18 Kremlyovskaya str., Kazan 420008, Russia}
\address{$^{3}$ Phys., Math. \& IT Dept., Baltic Fed. Univ., 14 A. Nevskogo str., Kaliningrad 236041, Russia}
\address{$^{4}$ Special Astrophysical Observatory RAS, Nizhnii Arkhyz, 369167, Russia}

\ead{ant.biryukov@gmail.com}

\begin{abstract}

Neutron stars are extremely strong cosmic magnets which fields are expected to decay with time. Here we report on the 
simple test of this process. Adopting a novel approach, we have estimated surface magnetic fields $B$ for 76 
radiopulsars (the most numerous subclass of the known isolated neutron stars) which ages $t$ are known independently. 
Focusing on the accurate evaluation of the precision of both quantities, we determined a significant power-law trend 
$B(t) \propto t^{-\beta}$ with index $\beta = 0.19^{+0.05}_{-0.06}$ at 95\%  C.L. The effects of the observational selection 
turn this value into the upper limit for the intrinsic field decay rate. If so, then neutron star crusts are 
close to the ``impurity-free crystals'', which results in a relatively slow magnetic fields decay. 
\end{abstract}

\section{Introduction}

The large-scale magnetic field is the key feature of the neutron stars (NS). Being as strong as $\sim 10^{8-14}$ Gs at 
the star's surface, it to a large extent controls many physical properties and observational manifestations of these 
compact objects: their rotational evolution, cooling, x-ray flux etc.

Magnetic field strength $B$ on the surface of an isolated NS unlikely remains constant, but is secularly decaying up to 
one order of magnitude throughout the entire lifetime of the star owing to the Ohmic decay of the electric currents 
located in the NS crust and/or core (e.g. \cite{agu08, vigano13})

In spite of the number of indirect evidences for the NS magnetic fields decay (by analyzing and reproducing of pulsar 
rotational and/or cooling parameters, e.g. \cite{pons07,anna12,ip14}) no direct observation of this process has 
been obtained so far. The main obstacle in this way is that ages and magnetic fields of individual {\it isolated} 
neutron stars are difficult to be estimated independently and 
with a confidently calculated precision  at the same time.

In our research we focus on the well-justified estimation of radiopulsars magnetic fields, ages and especially the 
uncertainties of the both quantities. Analyzing the apparent pulsars $B(t)$ distribution directly, we, therefore, 
provide a reliable constraint on the evolution of isolated NSs magnetic fields.

\section{Probing the magnetic field evolution}

Up to date, observed pulsar periods $P$ and their derivatives $\dot P$ remain the basic source for the massive and 
homogeneous magnetic fields estimations. The spin-down law $\dot P(B, P)$, which is assumed to be known from theory, 
provides a strict functional relationship between the timing parameters and the strength of the field induction $B$ on 
the star's surface. The general form of this law ($P\dot P \propto B^2$) is well fixed among theoreticians, while its 
details are dependent on the specific physical assumptions. However, recent achievements in numerical simulation of 
pulsar magnetospheres from ``first principles'' have shown that accurately calculated spin-down of a realistic isolated 
pulsar follows the simple equation 
\begin{equation}
    P\dot P = -\frac{4\pi^2 R^6(M)}{I(M) c^3}\cdot B^2 \cdot (k_0 + k_1 \sin^2\alpha),
    \label{eq:spitkovsky}
\end{equation}
where $\alpha$ is the magnetic obliquity (the angle between the spin and the magnetic axes of the star), $R$, $I$ and 
$M$ are the star radius, moment of inertia and mass, respectively, $c$ is the speed of light while $k_0 \approx 1$ and 
$k_1 \approx 1.4$ can be considered as universal constants for real  NSs \cite{spitkovsky06,phil15}. Dependencies $R(M)$ and $I(M)$ 
here are determined by the neutron star equation of state (EOS or eos hereafter).

The value of the 
logarithm\footnote{We adopt decimal logarithms within the paper} of the 
magnetic field extracted from this equation thus depends on the actual values of a neutron star mass, 
magnetic obliquity, spin period and its derivative and, finally, the equation of state.  Its estimation can be
obtained within a probabilistic approach by substituting the distributions of masses and 
magnetic angles of isolated pulsars to $\log B(P, \dot P, M, \alpha \mid {\rm EOS})$. As it has been shown in our previous research (\cite{bab17}, which we 
will hereafter refer to as BAB), the distribution of $\log B$ is nearly { Gaussian for
the representative subset of 22 EOSes with width $\sigma[\log B] \approx  0.06...0.09$ dex 
depending on the specific EOS, when state-of-the-art observational constraints to the pulsar masses and obliquities 
distributions are adopted (see BAB for details).

The value of $\sigma[\log B]$ being as small as $\sim 0.1$ dex represents the formal accuracy of the magnetic
field estimate and, finally, makes possible the accurate probing of the isolated NS magnetic fields evolution.

Another component which is required for this procedure is the knowledge of confident 
pulsars ages. Notice that the most common (and simple to calculate) so-called characteristic age
$\tau_{\rm ch} = P/(2\dot P)$ seems to be inapplicable here. Mostly because of its bias relative to the real ages of 
pulsars owing to nonzero initial spin period $P_0$. 

Therefore, we have compiled a list of 76 normal rotation-powered pulsars, whose ages were estimated independently of the 
$\tau_{\rm ch}$ and with  more or less confidently known uncertainties. The list consists of:
({\it i}) 22 normal pulsars associated with young (up to $10^5$ yr) supernova remnants (SNRs), ages of which have been 
discussed in the literature. This part is mostly the compilation of the data already prepared by other authors \cite{pt12, gill13} with some 
extensions. ({\it ii}) 36 pulsars with well-constrained kinematic ages $t_{\rm kin}$ mostly derived by Noutsos et al \cite{nout13}.
These data (along with $t_{\rm kin}$ uncertainties) have been extracted from the tracing of the galactic motion of 
particular objects back in time assuming that
they had been born within the galactic disk. ({\it iii}) 18 pulsars, whose kinematic ages have been derived with relatively low precision 
(all obtained in \cite{nout13}). 
These ones are quite old pulsars with typical $t_{\rm kin,0} \sim 3\times 10^7$ years. 

The empirical dependence of the calculated magnetic fields $\log B(P, \dot P)$ on the ages $\log t$ of the described 76 
pulsars is shown on the left plot of the figure 1. The values of the fields were calculated by substituting pulsar's $P$ and $\dot P$ to 
eq~(\ref{eq:spitkovsky})\footnote{Both parameters for every pulsar were taken from the ATNF database \cite{atnf} (version 1.54),\\ {\tt 
http://www.atnf.csiro.au/research/pulsar/psrcat/expert.html}}, while uncertainties of $\log B$ were modeled according to the
distribution of $\Delta^{\rm BSk21}_{\rm B}$ obtained in BAB.

The clear linear gradient of $\log B$ with an increase of $\mathbf{\log t}$ can be seen there: the estimated magnetic fields 
of aged pulsars in average are one order of magnitude smaller than that of young neutron stars.

To check the significance of this gradient we undertook a full Bayesian fit of the data by a minimal power law
$B(t) = B_0 \left( t/10^6\mbox{ years} \right)^{-\beta}$, where $B_{0}$ is the average field strength at age $10^6$ years, 
while $\beta$ is the field decay index. As it can be also seen from the plot, the scatter of $\log B$ values at a given age is larger than the typical 
uncertainty of the values of $\log B$. It implies the existence of the intrinsic scatter in the physical parameters of 
the neutron stars. To quantify it the variance of the scatter $\sigma^2_{\rm scatt}[\log B]$ has been also added in the Bayesian 
model.  All the calculations have been performed by means the {\it Just Another Gibbs Sampler}\footnote{\texttt{http://mcmc-jags.sourceforge.net/}}
package (JAGS, \cite{jags}), which uses the Markov chain Monte Carlo simulations for producing the posterior 
distributions of model parameters. Within the calculations, the formal errors of $\log B$ and $\log t_{\rm SNR/kin}$ have been taken into
account.

\begin{figure*}
\begin{center}
\includegraphics[width=7.3cm]{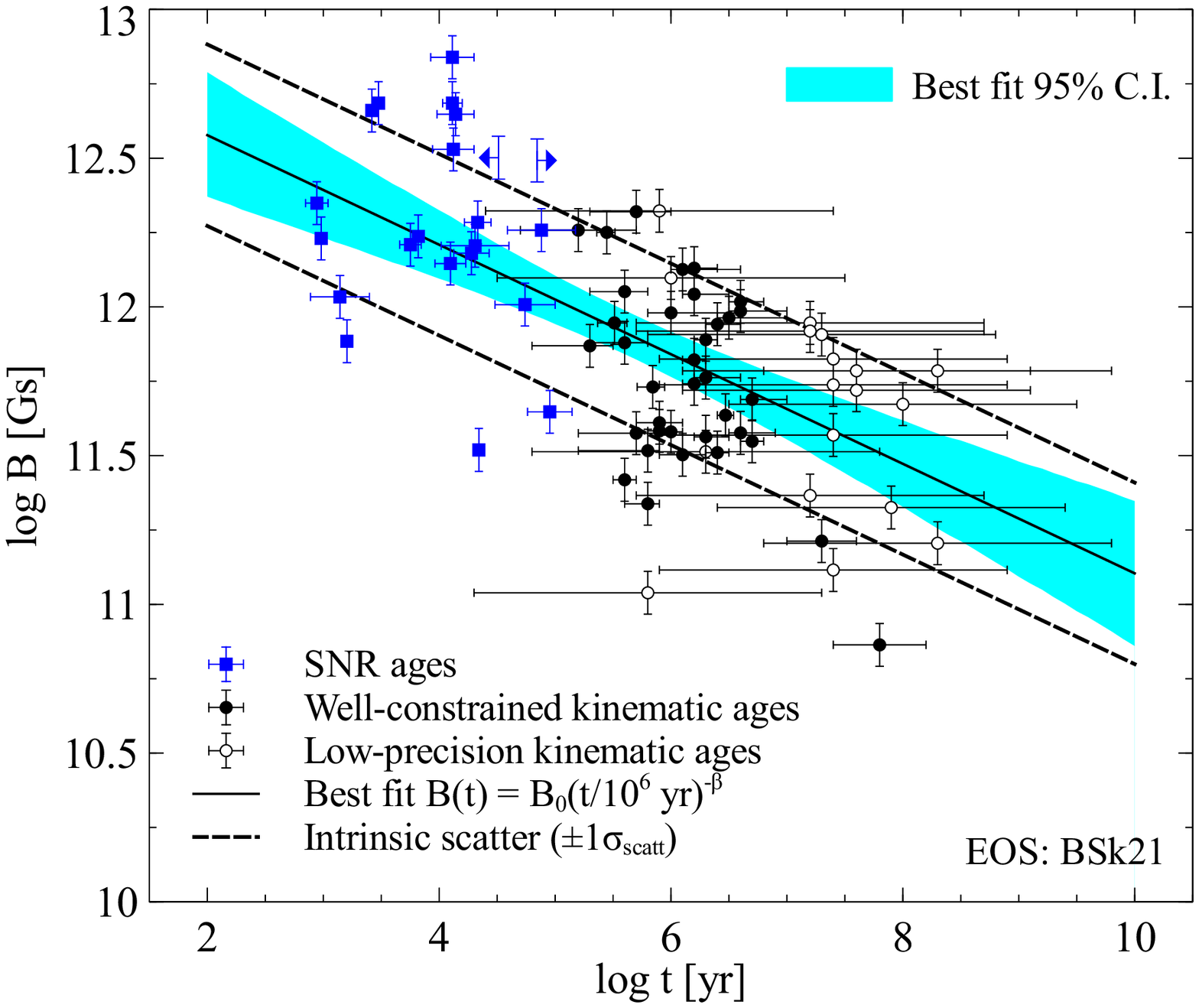}
\includegraphics[width=7.3cm]{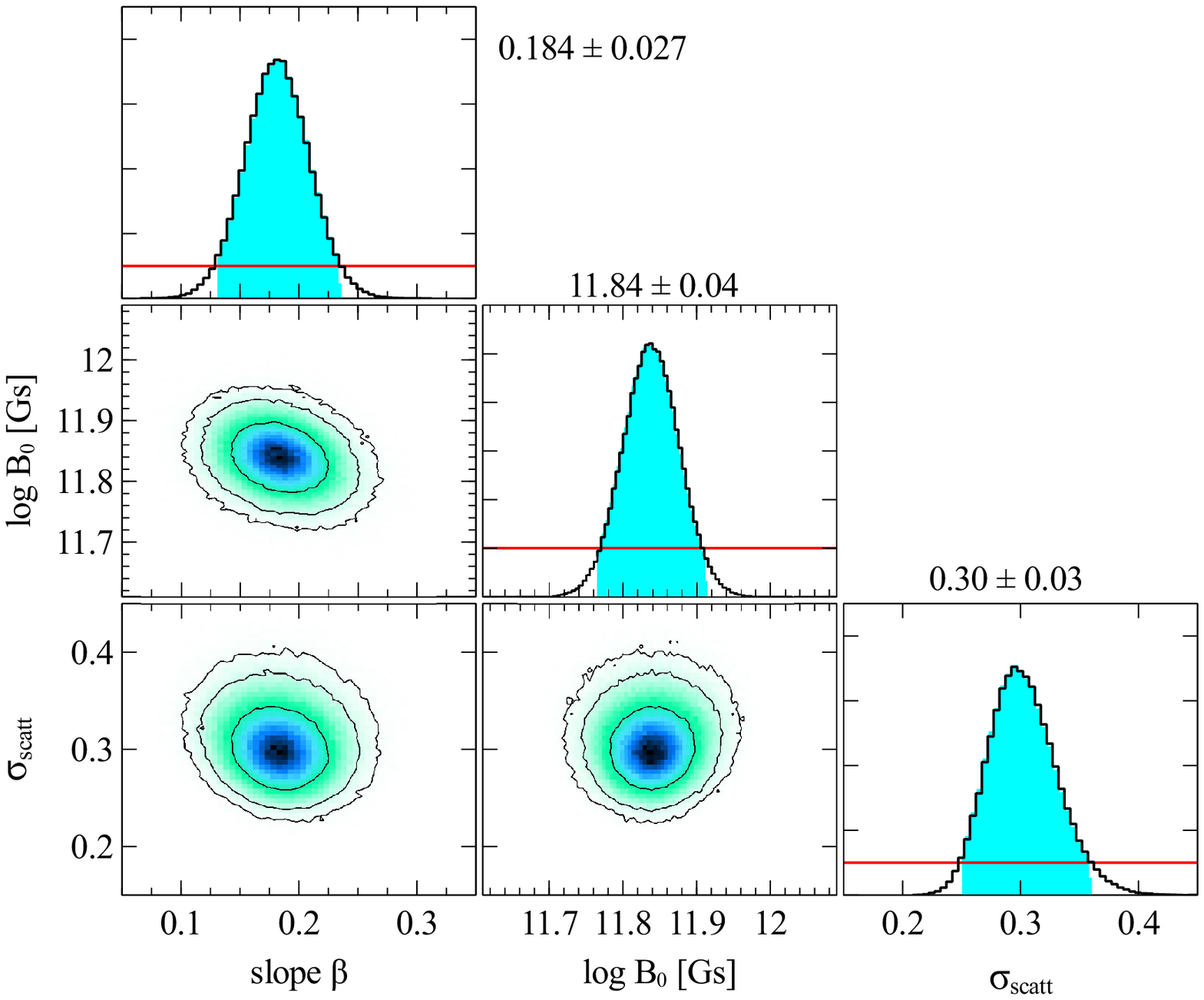}
\end{center}
\caption[]{{\it Left plot:} The apparent dependence of the timing-based magnetic field values,
calculated within the BSk21 equation of state, on the 
estimation of ages of 76 pulsars. The best fit by the power-law model is shown by 
the solid line with the corresponding 95 per cent  C.L. filled by cyan. The dashed lines
represent $\pm 1\sigma_{\rm scatt}[\log B]$ of estimated intrinsic scatter in the observed
$\log B$. The slope of the best fit $\beta \approx 1/5$ means the existence of a highly significant 
evolutionary trend in the magnetic fields of the considered pulsars. {\it Right plot:} The marginalized posterior 
distributions of the parameters of apparent pulsar magnetic fields evolution: the slope $\beta$, magnetic field $B_0$ 
at the age $10^6$ years and the intrinsic scatter standard deviation $\sigma_{\rm scatt}[\log B]$. The numbers given next to 
the plots are the averages and standard deviations for the
corresponding distributions. Uniform priors adopted for these parameters are shown by red lines on the same plots. The solid-line contours on the 2-d plots represent 68, 95 and 99\%  credible areas 
respectively.}
\end{figure*}

The results of the calculations are shown on the right plot of the figure 1, where the posterior 
distributions of the $\beta$, $\log B_0$ and $\sigma_{\rm scatt}[\log B]$ marginalized over individual pulsar parameters are 
plotted along with their correlations. The analysis of the apparent $\log B - \log t$ diagram shows that the magnetic fields of the considered pulsars
follow a highly significant trend so that $B(t) \propto t^{-1/5}$.

\section{Discussion and conclusions}

The provided fit to the collected data is nothing more than a minimal
phenomenological description of the apparent field evolution. The trend $B(t) \propto t^{-1/5}$ is likely to be 
significantly affected by the effect of the observational and analysis selection. Firstly, the systematic magnetic 
angle evolution mentioned above seems to be able to shift the estimated value of $\beta$. Indeed, average obliquity of 
older pulsars is expected to be less than that of younger ones. However, even within the most extremal case when {\it all}
young pulsars have magnetic angles $\alpha \approx 90^{\circ}$, which further evolve to $\alpha \approx 0^{\circ}$ at 
the time interval of $10^3\dots 10^8$ years, one can get the maximal bias in determined $\beta$ close to $\approx 0.04$.
This is comparable to the estimated accuracy of $\beta$ estimation. Hence, the obliquity evolution unlikely 
can affect the apparent slope of $\log B-\log t$ distribution.

At the same time we undertook an accurate population synthesis of normal isolated radiopulsars to figure out the main 
components of the observational selection which affect the apparent slope in $\log B - \log t$ diagram. We mostly
reproduced the model of pulsars evolution implemented in \cite{fgk06}, but adopting the Spitkovsky's spin-down equation 
(\ref{eq:spitkovsky}) along with the consistent model for the magnetic obliquity 
evolution \cite{phil15}.

As a result, it has been found, that even within the non-evolving magnetic fields, the distribution of 
``observed'' pulsars (i.e. those of them which successfully passed the instrumental selection) on the $\log B - \log t$ 
plane is significantly inclined with the slope $\beta \approx 0.3$. (Here we assume that modeled pulsar ages $t$ are unbaised relative to that estimated from 
PSR-SNR associations and NS motion within the Galaxy).

After a number of runs of the population synthesis (within different assumptions) we concluded 
that the obtained value $\beta \approx 0.2$ for the considered 76 pulsars represents {\it the upper limit} for the 
intrinsic rate of the magnetic field evolution. This means that our results likely reject possible rapid field decay, which, however, is predicted by a number of 
existing theoretical models.

\begin{figure*} 
\begin{center}
\includegraphics[width=15cm]{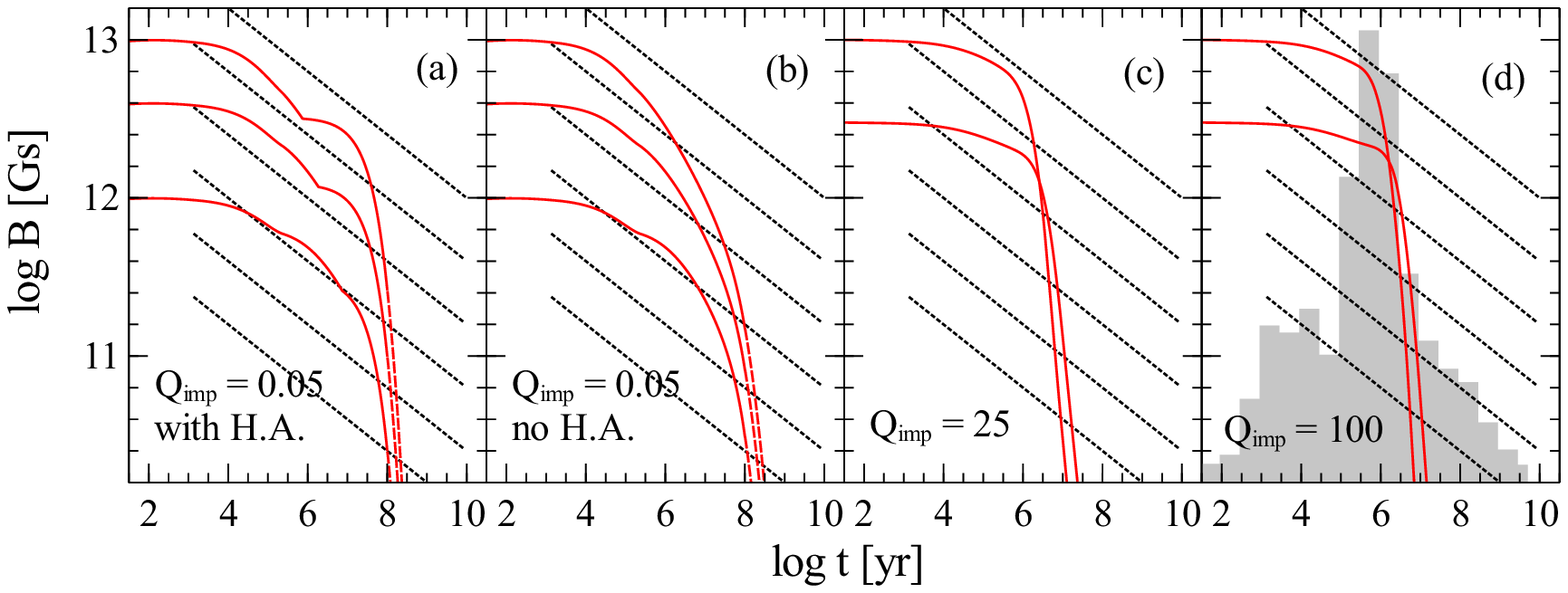}
\end{center}
\caption[]{Theoretical curves of pulsar magnetic fields decay (red solid 
lines) for different initial $B_{\rm init}$ and amount of impurities within a NS crust ($Q_{\rm imp}$ parameter). The dashed lines are plotted with the slope
$\beta = 0.2$ representing the most probable upper limit for the field decay rate obtained in the current work.
(a) The phenomenological model proposed in \cite{ip15}. (b) The same model as in the panel (a) but without 
the Hall attractor (H.A.); (c, d) The evolutionary tracks from \cite{vigano13} assuming 
larger amount of impurities. The empirical distribution of pulsar ages of the considered subset of 76 pulsars is also 
shown by the grey-filled histogram.}
\end{figure*}

 NS fields dissipation history can be roughly divided into two epochs. The first one lasts up to
$\sim 10^{4-6}$ years while the star's surface is hot enough and the electrons scattering from the phonons
dominates. The magnetic field changes relatively weakly within this stage.
In contrast, during the second epoch, the electrons scattering from the impurities of
the NS crust crystalline lattice become important and can significantly increase the rate of the
field decay. The amount of the impurities within the crust matter 
can be described by the parameter $Q_{\rm imp}$, which
higher values lead to the faster magnetic field decay of middle-age and old pulsars \cite{shter06}. Generally, there is no consensus about the value of 
$Q_{\rm imp}$ for the realistic neutron 
stars. Although the crusts of normal pulsars  are likely to be close to
``impurity-free crystals'' ($Q_{\rm imp} \sim 0$) in contrast to that of the magnetars \cite{vigano13, gullon14}. 

In figure 2 four models of the pulsars magnetic field decay are overplotted on the 
lines representing the  upper limit for the decay rate consistent with our results.\footnote
{We are grateful to Andrey Igoshev and Miguel Gull{\'o}n for providing us with these theoretical curves of the magnetic 
field decay.}

It is clearly seen that the models proposed in \cite{ip15}, assuming the low $Q_{\rm imp}$, are much closer to 
the weak field decay scenario, which satisfy the empirical
data definitely better. Additionally notice that rapidly decaying magnetic field makes it hard to explain the 
observation of active pulsars with ages greater than $\sim 10^6$ years which exist in the considered subset (see figure 2d).

Thus, the final results of our research can be summarized as follows: ({\it i}) Using the ages of 22 supernova remnants associated with young pulsars and 54 
kinematic ages of older sources as 
well as the refined version of the timing-based estimator of pulsars magnetic fields, we have found significant 
apparent trend $B(t)$ in the magnetic fields of normal radiopulsars so that $B(t) \propto t^{-1/5}$ 
({\it ii}) The ascertained trend $B(t)$ is 
significantly affected by the effects of the observational selections. The accurate
population synthesis of the observed pulsars has shown that the slope $\beta$ in the apparent $\log B - \log t$
distribution is likely to be the {\it upper limit} of the intrinsic one $\beta_0$. ({\it iii}) Thus, our result confirms that crusts of normal (rotation-powered) 
pulsars are likely to be close to ``impurity-free crystals'', which lead to a relatively slow surface magnetic field decay at the final stages of their 
active life. At the same time we reject the theoretical models of NS magnetic fields evolution that would assume high amount of nuclear impurities within their 
crusts.

 \ack
{ We are grateful to Andrei Igoshev, Sergey Popov, Stefano Andreon, Yury Lyubarsky and J$\acute{\rm e}$r$\hat{\rm o}$me P$\acute{\rm e}$tri for their comments and fruitful discussions.The work is performed according to the Russian Government Program of Competitive
Growth of Kazan Federal University. Data analysis and simulations used hardware and software supported by
the Russian Science Foundation grant No. 14-50-00043. Artyom Astashenok thanks Russian Ministry of Education and 
Science for the support 
(project 2058/60).}

\section*{References}

\providecommand{\newblock}{}

\end{document}